\documentclass[journal,12pt,draftclsnofoot,onecolumn]{IEEEtran}
\IEEEoverridecommandlockouts
\ifCLASSINFOpdf
\else
\fi
\newtheorem{theorem}{Theorem}
\usepackage{amsmath,graphicx,cite}
\begin{document}
\begin{titlepage}
\begin{center}
\vspace*{-2\baselineskip}
\begin{minipage}[l]{7cm}
\flushleft
\includegraphics[width=2 in]{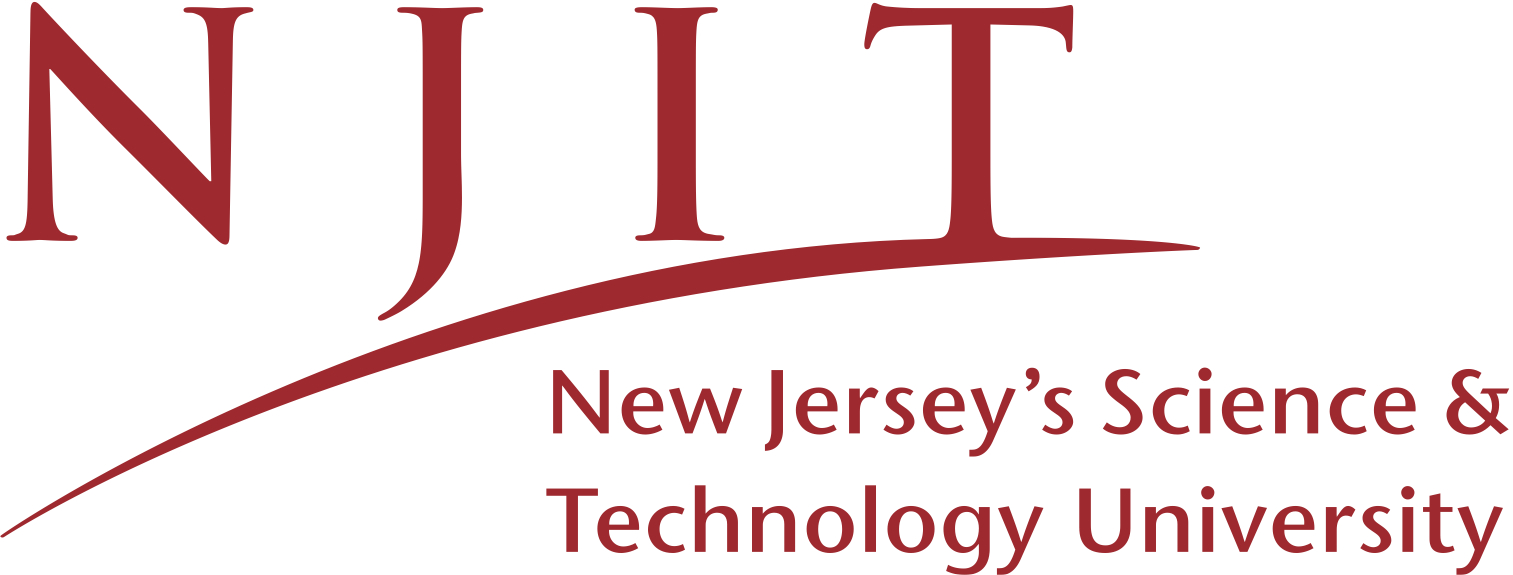}
\end{minipage}
\hfill
\begin{minipage}[r]{7cm}
\flushright
\includegraphics[width=1 in]{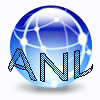}
\end{minipage}

\vfill

\textsc{\LARGE Green Energy Aware Avatar Migration Strategy in Green Cloudlet Networks\\[12pt]}
\vfill
\textsc{%\LARGE Author Name\\[12pt]
\LARGE  XIANG SUN \\ NIRWAN ANSARI\\QIANG FAN}\\
\vfill
\textsc{\LARGE TR-ANL-2015-006\\[12pt]
\LARGE Sep 07, 2015}\\[1.5cm]
% Bottom of the page
\vfill
{ADVANCED NETWORKING LABORATORY\\
 DEPARTMENT OF ELECTRICAL AND COMPUTER ENGINEERING\\
 NEW JERSY INSTITUTE OF TECHNOLOGY}
\end{center}
\end{titlepage}

\title{Green Energy Aware Avatar Migration Strategy in Green Cloudlet Networks}
\author{\IEEEauthorblockN{Xiang Sun,~\IEEEmembership{Member,~IEEE}, Nirwan Ansari,~\IEEEmembership{Fellow,~IEEE}, and Qiang Fan,~\IEEEmembership{Member,~IEEE}}
\thanks{This work was supported in part by the National Science Foundation under grant no. CNS-1320468. Xiang Sun, Nirwan Ansari and Qiang Fan are with the Advanced Networking Laboratory, Department of Electrical and Computer Engineering, New Jersey Institute of Technology, Newark, NJ, 07102, USA. (Email: {xs47, nirwan.ansari, qf4}@njit.edu).}
}
\maketitle

\section{abstract}
We propose a Green Cloudlet Network (\emph{GCN}) architecture to provide seamless Mobile Cloud Computing (\emph{MCC}) services to User Equipments (\emph{UE}s) with low latency in which each cloudlet is powered by both green and brown energy. Fully utilizing green energy can significantly reduce the operational cost of cloudlet providers. However, owing to the spatial dynamics of energy demand and green energy generation, the energy gap among different cloudlets in the network is unbalanced, i.e., some cloudlets' energy demands can be fully provided by green energy but others need to utilize on-grid energy (i.e., brown energy) to satisfy their energy demands. We propose a Green-energy awarE Avatar migRation (\emph{GEAR}) strategy to minimize the on-grid energy consumption in GCN by redistributing the energy demands via Avatar migration among cloudlets according to cloudlets' green energy generation. Furthermore, GEAR ensures the Service Level Agreement (\emph{SLA}) in terms of the maximum Avatar propagation delay by avoiding Avatars hosted in the remote cloudlets. We formulate the GEAR strategy as a mixed integer linear programming problem, which is NP-hard, and thus apply the Branch and Bound search to find its sub-optimal solution. Simulation results demonstrate that GEAR can save on-grid energy consumption significantly as compared to the Follow me AvataR (\emph{FAR}) migration strategy, which aims to minimize the propagation delay between an UE and its Avatar.

\section{Keywords}
Mobile cloud computing, cloudlet, live migration, energy optimization, Branch and Bound search.\

\section{Introduction}
The emergence of Mobile Cloud Computing (\emph{MCC}) is enabling execution of computation-intensive applications (e.g., augmented reality and speech recognition) in a user equipment (\emph{UE}), i.e., the UE can offload some tasks to high performance Virtual Machines (\emph{VM}s) in a data center and VMs can help the UE execute these tasks in order to improve the task execution time and reduce the UE's energy consumption. However, the existing MCC architecture suffers from the long communications latency between a UE and its VM in a remote data center as the communications link traverses the Wide Area Network (\emph{WAN}) which does not guarantee any minimum QoS to the UE; it is also very hard to control the WAN latency \cite{1}. According to a report \cite{2}, “Amazon famously claimed that every 100 millisecond reduction in delay led to a one percent increase in sales. Google also stated that for every half second delay, it saw a 20 percent reduction in traffic.” Therefore, reducing the latency can bring a huge benefit to the application providers. The concept of cloudlets has thus been proposed to reduce the propagation delay between a UE and its VM \cite{1}. A cloudlet is a tiny version of the data center and is located close to the UE, and so communications between the UE and its VM can be established via the local area network (\emph{LAN}).\
\begin{figure}[!htb]
	\centering	
	\includegraphics[width=1.0\columnwidth]{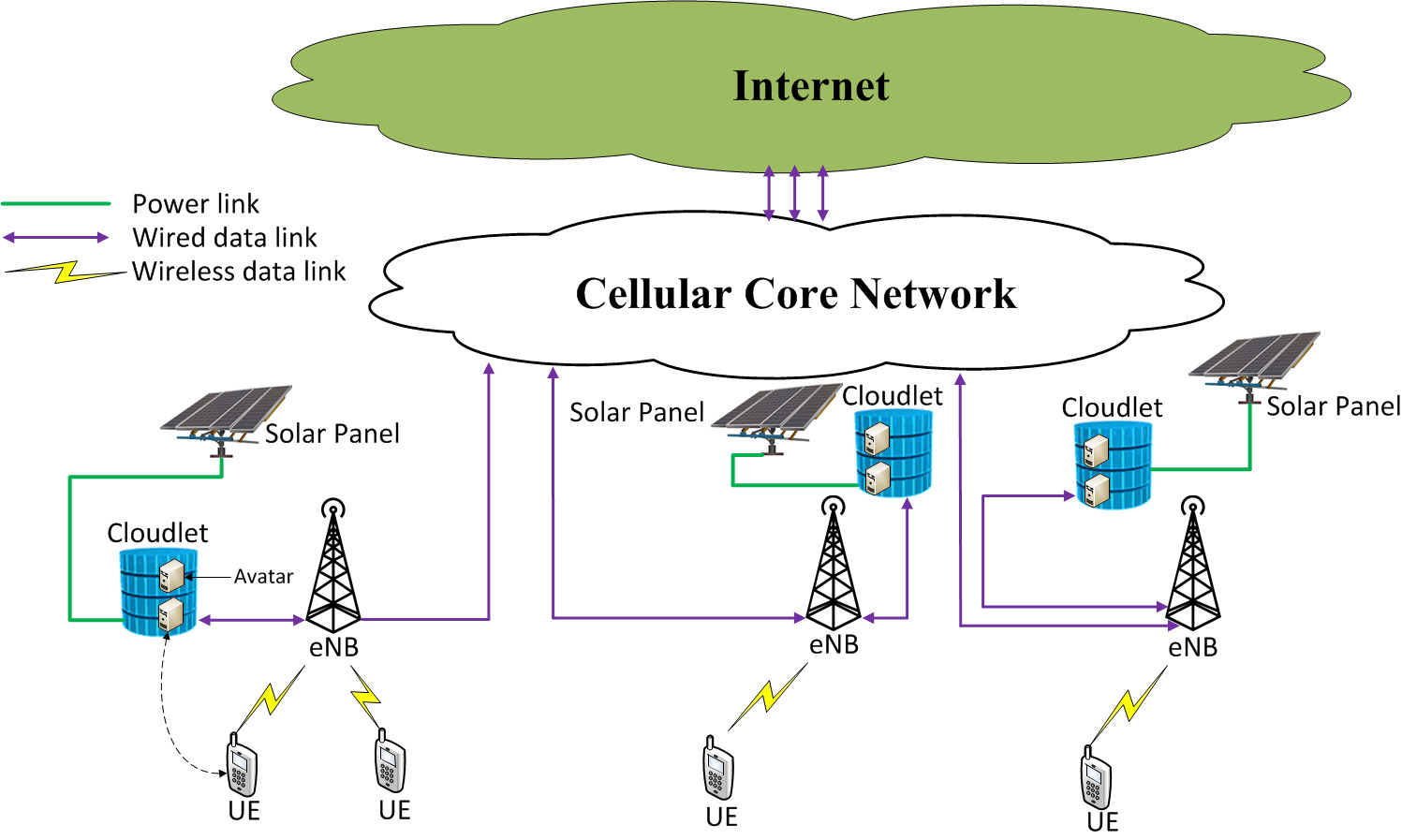}
	\caption{Green cloudlet network architecture.}	
	\label{fig1}
\end{figure}

To reap benefits of cloudlets and make them sustainable, we propose the Green Cloudlet Network (\emph{GCN}) architecture as shown in Fig. \ref{fig1}. Since the existing LTE network infrastructure can provide seamless connection between a UE and its eNB, each eNB is connected with a cloudlet via high speed fiber so that UEs can utilize the MCC technology everywhere. Meanwhile, the cloudlet is so close to the UE that the propagation delay is minimized. Each MCC UE subscribes one Avatar, a high performance VM in the cloudlet, to help run different tasks and provide extra storage space. Avatars are software clones of their UEs and always available to UEs when UEs are moving from one area to the others \cite{3}. Moreover, in order to overcome the inefficient structure of the traditional Evolved Packet Core (\emph{EPC}) network (i.e., the control plane and the data plane are centralized in the Packet data network GateWay (\emph{P-GW}) and Serving GateWay (\emph{S-GW}) \cite{4}),  Software Defined Network (\emph{SDN}) based cellular core network \cite{5} has been proposed in the GCN architecture in order to decouple the control plane (which is implemented in the SDN controller) and data plane (which is implemented in openflow switches) and provide  efficient and flexible communications paths between Avatars in different cloudlets or between UEs in different eNBs.  \

\begin{figure}[!htb]
	\centering	
	\includegraphics[width=1.0\columnwidth]{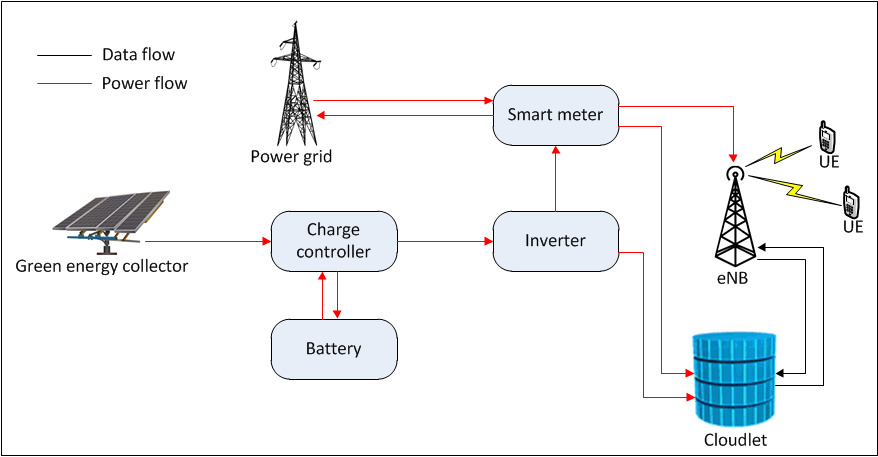}
	\caption{The power supply system of the cloudlet.}
	
	\label{fig2}
\end{figure}

GCN facilitates communications between a UE and its Avatar, but the distributed cloudlets increase the operational cost of cloudlet providers and $CO_{2}$ emission, i.e., a huge amount of on-grid energy (we assume energy from the power grid is brown energy, and thus the terms on-grid energy and brown energy are interchangeable for the rest of paper) will be consumed in order to maintain the GCN infrastructure. In order to reduce on-grid energy, “greening” is introduced in the GCN architecture, i.e., each cloudlet is powered by green energy generated from solar panels or other green energy collectors and uses on-grid energy as a backup. The power supply system of each cloudlet is shown in Fig. \ref{fig2}, in which the green energy collector absorbs energy from the green energy source (i.e., solar radiation) and converts it into electrical power, the charge controller regulates the electrical power from the green energy collector, and the electrical power is converted between AC and DC by the inverters. The smart meter records the electric energy from the power grid consumed by the cloudlet and eNB. Note that green energy in Fig. \ref{fig2} is only available to the cloudlet in this system. Certainly, eNBs can be equipped with its own green energy supply system as well, but green energy from the same green energy collector cannot be shared by the cloudlet and the eNB simultaneously (because the cloudlet provider and the LTE network provider play different roles in the network. Thereby, they would not share the infrastructure with each other). Thus, green energy supplements of the cloudlet and the eNB are independent with each other. In this paper, we only consider how to efficiently utilize green energy by cloudlets in the network.  \

Owing to the spatial dynamics of the distribution of UEs among different eNBs' coverage areas and the dynamics of application loads among Avatars, different cloudlets may require different energy demands for running the application loads of the hosting Avatars. Meanwhile, green energy generation also exhibits spatial dynamics. Therefore, some cloudlets, which have less energy demand and more green energy generated, would have excess of green energy. Conversely, some cloudlets, which have more energy demand and less green energy generated, would pull energy from the power grid (non-renewal). Such unbalanced energy gap (energy demand minus green energy generation) among different cloudlets increases the on-grid energy consumption. Therefore, fully utilizing green energy can tremendously reduce the on-grid energy consumption, and thus potentially decreases the operational cost of the cloudlet providers and $CO_{2}$ emission. In this paper, we propose the Green-energy awarE Avatar migRation (\emph{GEAR}) strategy to minimize the on-grid energy consumption in GCN by redistributing the energy demands in terms of migrating Avatars among cloudlets according to cloudlets' green energy generation. Meanwhile, GEAR also guarantees the SLA, which is defined as the maximum Avatar propagation delay, i.e., the maximum propagation delay between the UE's eNB (the eNB which is serving the UE) and the UE's Avatar (the propagation delay between the UE's eNB and UE's Avatar may become higher over time, because the UE may move away from the original place or UE's Avatar may migrate to the cloudlet which is far away from the UE, and so the propagation delay may become humongous).\

The rest of the paper is organized as follows. In Section II, we briefly review the related works. In Section III, we setup a power consumption model of a cloudlet. In Section IV, we formulate the avatar migration strategy in order to minimize the on-grid energy consumption. In Section V, we demonstrate the viability of GEAR via simulation results. The conclusion is presented in Section VI. 

\section{Related Works}
Previous works \cite{1}\cite{6} have proved that cloudlets can significantly reduce the communications latency between UEs and VMs in the cloudlet. Tarik and Ksentini \cite{7} introduced a follow-me cloud, i.e., a UE's service is continuously migrated to the data center which is much closer to the UE. Follow-Me Cloud tries to minimize the propagation delay between a UE and its VM in the data center, but it does not capitalize on green energy to optimize the energy consumption. \

Rather than considering the communications latency between a UE and its VM, many works have focused on optimizing energy usage in Internet-scale Data Centers (\emph{IDC}s). Studies \cite{8}\cite{9}\cite{10}\cite{11}\cite{12} have aimed to minimize the electricity cost in IDCs which are only powered by brown energy, i.e., the workloads are migrated from high electricity cost IDCs to low electricity cost IDCs. Gkatzikis and Koutsopoulos \cite{13} showed that introducing green energy in the cloud can significantly reduce the usage of brown energy, but there is a big challenge to match the dynamic green energy generation and dynamic energy demands of data centers in the green cloud network. Hatzopoulos \emph{et al.} \cite{14} assumed each task request is assigned to one VM and they tried to allocate running VMs into different data centers so that the total cost of power consumption in the green cloud network is reduced and the deadline of each request's response time is ensured. By considering the daily/seasonal effects of the green energy supplement in each geographical data center, Chen \emph{et al.} \cite{15} proposed a holistic workload scheduling algorithm in order to minimize the brown energy consumption across all the data centers. Studies \cite{16}\cite{17} proposed the similar idea. Both of them design a profit maximization strategy which is to assign the incoming workloads among geo-distributed green data centers by considering the price of electricity, renewable power generation and SLA parameters.\

As compared to the previous efforts, this paper presents several enhancements. First, GCN is implemented in order to provide seamless MCC services to UEs with lower and controllable communications latency. Different from traditional energy optimization in the green cloud network, SLA is defined as the maximum Avatar propagation delay (i.e., the propagation delay bound between a UE's eNB and UE's Avatar), and so our objective is not only to minimize the on-grid energy but also to guarantee the predefined SLA for each UE in GCN. Ensuring the SLA is important for cloudlet providers, because each UE is moving over time and UEs' Avatars may migrate to the cloudlet with higher green energy generation and lower energy demands. Thus, some UEs' Avatars may be far away from themselves leading to the high propagation delay. As mentioned previously, the unbearable latency degrades the performance of MCC applications. Thus, guaranteeing the maximum propagation delay for each UE is an important factor to be considered when we design an optimal Avatar migration strategy. To the best of our knowledge, existing literature has not addressed this issue. Second, we propose a novel live Avatar migration strategy, i.e., GEAR, to achieve our objectives, and demonstrate the reduction of on-grid energy consumption without violating the SLA via extensive simulations.

\section{System Model}

Cloudlets are distributed in the network powered by both on-grid and green energy. We assume that the servers in GCN are homogeneous, i.e., the configuration of each server is the same. Every UE's Avatar is also homogeneous, but the application loads of different UEs' Avatars are different. So, each server can host a fixed number of Avatars $\tau$, but the application loads on different servers vary. The cloudlet provider plays the role of an Infrastructure as a Service (\emph{IaaS}) provider, i.e., the cloudlet provider supplies virtualized computing resources in terms of Avatars to UEs. Although the provider does not impose any SLA of applications onto Avatars, it does ensure the SLA to UEs, i.e., the maximum Avatar propagation delay $\varepsilon$.\

As mentioned earlier, Avatar migration is enabled to adjust the energy demand among the cloudlets. However, live migration may cost several seconds or minutes to activate an Avatar moving from the source server in one cloudlet to the destination server in another cloudlet, and so proactive migration decision should be made, i.e., the decision maker should determine each Avatar's location (i.e., in which cloudlet) in the next time slot based on the prediction of the energy demand and the green energy generation of each cloudlet. The hourly solar energy generation can be estimated by using typical annual meteorological weather data for a given geolocation \cite{18}. Meanwhile, the estimation of the cloudlet's energy demand can be calculated by means of forecasting the power consumption of each active server in the cloudlet as will be discussed later by setting up the active server power consumption model. \

In order to identify the location of different Avatars, two indicator functions $\delta(i,k)$ and $\eta_i(j,k)$ are introduced where $i$ is the index of cloudlets in the network, $j$ is the index of servers in one cloudlet, and $k$ is the index of Avatars in the network. So, $\delta(i,k)=1$ implies that Avatar $k$ is located in cloudlet $i$. Meanwhile, $\eta_i(j,k)=1$ indicates that Avatar $k$ is in cloudlet $i$'s server $j$. Therefore,
\begin{equation}
\delta \left( i,k \right)=\sum\limits_{j=1}^{{{n}_{i}}}{{{\eta }_{i}}\left( j,k \right)}														
\end{equation}
where $n_i$ is the number of active servers in cloudlet $i$ and it is a function of $\delta(i,k)$:

\begin{equation}
{{n}_{i}}=\left\lceil \frac{\sum\limits_{k}{\delta \left( i,k \right)}}{\tau } \right\rceil 
\end{equation}
where $\tau$ is the number of Avatars hosted in the server.
	\subsection{Server Power Consumption Model}
		In this section, we model the power consumption of active servers in a cloudlet. The power consumption of active server $j$ in cloudlet $i$ can be characterized as follows \cite{19}:
		\begin{equation}
			{{P}_{i,j}}={{P}^{s}}+P_{i,j}^{vir}+\alpha \times u_{i,j}^{app}
		\end{equation}		
where $P_{i,j}$ is the total power consumption of active server $j$ in cloudlet $i$; $P^s$ is the power consumption of the server when it is in the standby mode, i.e., when the server's CPU load is zero; $P_{i,j}^{vir}$ is the power consumption of server $j$ for doing virtualization and we will discuss it in the next paragraph; $\alpha_{i,j}^{app}$ is the power consumption of server $j$ for running different applications on its hosting Avatars, where $u_{i,j}^{app}$ is server $j$'s CPU usage for running Avatar $i$'s application load and $\alpha$ is the coefficient that maps the CPU usage into the power consumption. So, if the CPU usage for running Avatar $k$'s application load is $u_{k}^{app}$, then the server $j$'s CPU usage can be considered as a function of $\eta_i (j,k)$:
		\begin{equation}
			u_{i,j}^{app}=\sum\limits_{k}{{{\eta }_{i}}\left( j,k \right)\times u_{k}^{app}}
		\end{equation}	
			
As mentioned earlier, $P_{i,j}^{vir}$ is the power consumption of server $j$ for performing virtualization, and it includes two parts [19]:		
		\begin{equation}
			P_{i,j}^{vir}=P_{i,j}^{hyper}+P_{i,j}^{idle}
		\end{equation}		
where $P_{i,j}^{hyper}$ is the power consumption of server $j$ in cloudlet $i$ for running a hypervisor (i.e., a VM manager) without any Avatar load (i.e., the hypervisor only manages the configuration of different Avatars in the server). In order to determine the power consumption of an idle hypervisor, Warkozek \emph{et al.} \cite{19} showed that $P_{i,j}^{hyper}$ is proportional to the number of Avatars hosted in the server, i.e., the more Avatars server $j$ hosts, the more power is consumed on the hypervisor for configuring these Avatars:		
		\begin{equation}
			P_{i,j}^{hyper}=\beta \times \sum\limits_{k}{{{\eta }_{i}}\left( j,k \right)}
		\end{equation}		
where $\sum\nolimits_{k}{{{\eta }_{i}}(j,k)}$ indicates the total number of Avatars in sever $j$ and $\beta$ is the Avatar power coefficient, which is the power cost of the hypervisor for maintaining one Avatar.\
$P_{i,j}^{idle}$ in Eq. (5) is the power consumption of Avatars in the idle mode (Avatar does not take any application load from UE, but runs basic system operation instances) in server $j$ of cloudlet $i$. $P_{i,j}^{idle}$ is determined by the number of Avatars in server $j$ and the amount of CPU usage for running the operating system (\emph{OS}) kernel instances for each idle Avatar. Assuming that all UEs' Avatars use the same OS, thereby CPU usage for running the OS kernel instances $u^{idle}$ is the same for all Avatars. Therefore, we have
		\begin{equation}
		P_{i,j}^{idle}=\alpha \times {{u}^{idle}}\times \sum\limits_{k}{{{\eta }_{i}}\left( j,k \right)}
		\end{equation}
		
Substituting Eqs. (4)-(7) into Eq. (3) yields the power consumption of server $j$ in cloudlet $i$ as follows:		 
		\begin{equation}
		{{P}_{i,j}}={{P}^{s}}+\beta \times \sum\limits_{k}{{{\eta }_{i}}\left( j,k \right)}+\sum\limits_{k}{\left[ {{\eta }_{i}}\left( j,k \right)\times \alpha \left( {{u}^{idle}}+u_{k}^{app} \right) \right]}
		\end{equation}
		
		Since the servers in the cloudlet network are homogeneous, $P^s$ and $\alpha$, which are constants, can be pre-determined. Meanwhile, if all servers are installed with the same type of hypervisor, such as Hyper-V, ESX or Xen, then $\beta$ is the same for all servers. We define ${{u}_{k}}={{u}^{idle}}+u_{k}^{app}$ as the total CPU usage (including OS kernel CPU usage and application load CPU usage) for running Avatar $k$ in the server, and so Eq. (8) can be expressed as:
		\begin{equation}
			{{P}_{i,j}}={{P}^{s}}+\beta \times \sum\limits_{k}{{{\eta }_{i}}\left( j,k \right)}+\sum\limits_{k}{\left[ {{\eta }_{i}}\left( j,k \right)\times \alpha {{u}_{k}} \right]}
		\end{equation}
	\subsection{Cloudlet Power Consumption Model}
		Aside from running the cooling system and cloudlet network equipment, the major power consumption of a data center is the power consumed by the active servers. However, a cloudlet is a tiny version of the data center that does not need to maintain a powerful cooling system and plenty of switches, and so we assume all the power consumption of a cloudlet is contributed by the computing equipments such as servers, and we calculate cloudlet $i$'s power consumption as the sum of the power consumption of the active servers:
		\begin{equation}
			{{P}_{i}}=\sum\limits_{j=1}^{{{n}_{i}}}{{{P}_{i,j}}}={{n}_{i}}{{P}^{s}}+\beta \times \sum\limits_{j=1}^{{{n}_{i}}}{\sum\limits_{k}{{{\eta }_{i}}\left( j,k \right)}+\sum\limits_{j=1}^{{{n}_{i}}}{\sum\limits_{k}{\left[ {{\eta }_{i}}\left( j,k \right)\times \alpha {{u}_{k}} \right]}}}
		\end{equation}
		By approximating Eq. (2) into ${{n}_{i}}\approx \frac{\sum\nolimits_{k}{\delta (i,k)}}{\tau }$ and substituting Eq. (1) and ${{n}_{i}}\approx \frac{\sum\nolimits_{k}{\delta (i,k)}}{\tau }$into Eq. (10), we have the power consumption of cloudlet $i$:		
		\begin{equation}
			{{P}_{i}}\approx \sum\limits_{k}{\left[ \delta \left( i,k \right)\times \left( \frac{{{P}_{s}}}{\tau }+\beta +\alpha {{u}_{k}} \right) \right]}
		\end{equation}		
	\subsection{Avatar Propagation Delay Model}
	\begin{figure}[!htb]
		\centering	
		\includegraphics[width=1.0\columnwidth]{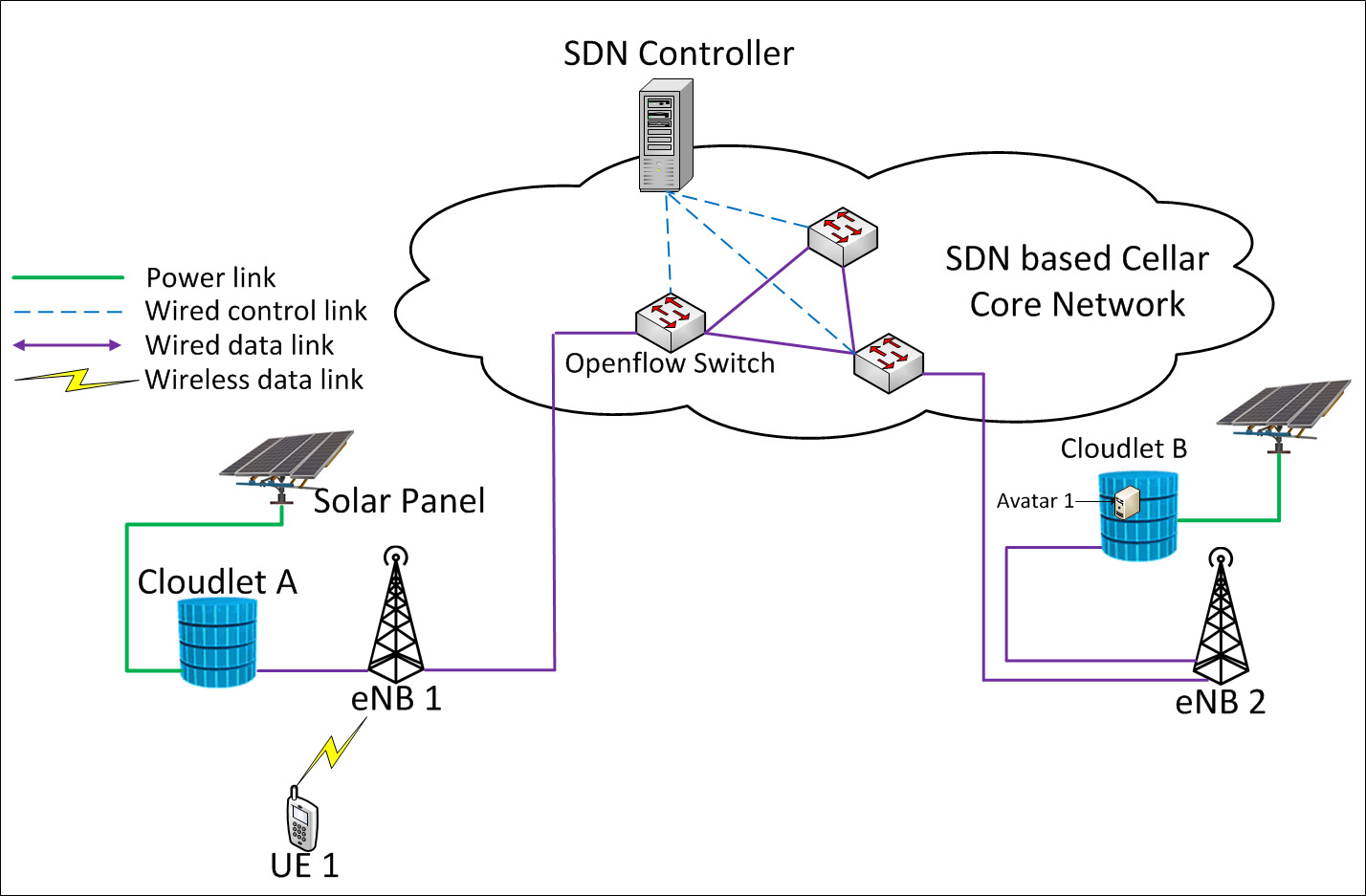}
		\caption{The communications between a UE and its Avatar}
		
		\label{fig3}
	\end{figure}
	Usually, a UE and its Avatar may not associate with the same eNB. As mentioned previously, UEs are moving over time and Avatars migrate to the cloudlet with more green energy generation and less energy demands. Thus, the communications between a UE and its Avatar might traverse SDN based cellular core network. As shown in Fig. \ref{fig3}, if UE 1 tries to communicate with its Avatar (i.e., Avatar 1 in cloudlet $B$), the communications path should traverse eNB 1, openflow switches, eNB 2 and cloudlet $B$. Thus, the communications delay between a UE and its Avatar comprises three parts: wireless communications delay between the UE and its eNB, propagation delay between the eNB and the cloudlet where the UE's Avatar is located, and the propagation delay within the cloudlet. However, the wireless communications delay is determined by the UE's billing plan and the LTE network provider's bandwidth allocation strategy which are not controlled by the cloudlet provider. Meanwhile, we assume the propagation delay within the cloudlet is negligible. So, we define Avatar propagation delay as the latency for propagating one packet between the UE's eNB and its cloudlet. Cloudlet providers only need to guarantee the SLA in terms of maximum Avatar propagation delay for each UE.\
		
	If the UE's Avatar is located at cloudlet $i$ and the UE is associated with eNB $e$ ($e$ is the index of eNB in GCN), we can express the Avatar propagation delay as $T_{i,e}$, which comprises two parts: $T_{i,e}^{prop}$, i.e., the propagation delay for transmitting a packet between cloudlet $i$ in which the UE's Avatar resides and the UE's eNB $e$; $T_{i,e}^{proc}$, i.e., the total processing delay for all the openflow switches on the routing path in processing one packet. Assume $T_{i,e}^{prop}$ is proportional to the distance $d_{i,e}$ \cite{20}, i.e., the distance between cloudlet $i$ in which the UE's Avatar resides and the UE's eNB $e$, i.e., $T_{i,e}^{prop}\propto {{d}_{i,e}}$ . Meanwhile, if we assume that the number of openflow switches on the routing path is also proportional to $d_{i,e}$ (i.e., the longer distance between Avatar's cloudlet and the UE's eNB, the more openflow switches the packet needs to traverse) and the average packet processing time on every openflow switch is the same, then $T_{i,e}^{proc}$ is also proportional to $d_{i,e}$, i.e., $T_{i,e}^{proc}\propto {{d}_{i,e}}$ . So, we conclude that:
	\begin{equation}
		{{T}_{i,e}}=\sigma \times {{d}_{i,e}}\times \delta \left( i,k \right)
	\end{equation}
	where $\sigma$ is the coefficient that maps the distance to the time delay. Assume the locations of each eNB and its cloudlet are known; note that Avatar is tracking the location of its UE all the time. Therefore, given the locations of each eNB and cloudlet, $d_{i,e}$ is a constant. Thereby, Avatar propagation delay is a function of $\delta(i,k)$.

\section{Problem Formulation}	
	Owing to the spatial dynamics of energy demand and green energy generation among different cloudlets, energy demand of some cloudlets can be met by green energy, but some cannot and need to consume on-grid energy. Meanwhile, owing to the disadvantages of “banking” green energy \cite{21}	, we assume that green energy is disinclined to be stored for each cloudlet \cite{22}. Therefore, green energy should be fully utilized in each time slot so that the on-gird energy can be minimized. Denote $\rho_i$ as the on-grid energy consumption of cloudlet $i$, i.e., ${{\rho }_{i}}=\max [0,\Delta T({{P}_{i}}-{{G}_{i}})]$, where $\Delta T$ is the length of one time slot; $P_i$ and $G_i$ are the power demand and the green power generation of cloudlet $i$. The objective of the GEAR strategy is to minimize the on-grid energy consumption of GCN in each time slot. So, we formulate GEAR as follows:	
	\begin{align}
	& \underset{\delta \left( i,k \right)}{\mathop{\min }}\,\text{   }\sum\limits_{i}{{{\rho }_{i}}} \\ 
	& s.t.\text{    }\forall i,\text{  }{{\rho }_{i}}\ge \sum\limits_{k}{\left[ \delta \left( i,k \right)\times \left( \frac{{{P}_{s}}}{\tau }+\beta +\alpha {{u}_{k}} \right) \right]}-{{G}_{i}}, \\ 
	& \forall i,\text{   }{{\rho }_{i}}\ge 0, \\ 
	& \forall k,\text{  }\sigma {{d}_{i,e}}\delta \left( i,k \right)\le \varepsilon , \\ 
	& \forall i,\text{   }\frac{1}{\tau }\text{ }\sum\limits_{k}{\delta \left( i,k \right)}\le {{m}_{i}}, \\ 
	& \forall k,\text{   }\sum\limits_{i}{\delta \left( i,k \right)}=1, 
	\end{align}	
	where  $\delta(i,k)$ is a binary variable indicating the location of Avatars, $\epsilon$ is the SLA provided by the cloudlet provider, and $m_i$ is the capacity of cloudlet $i$, i.e., the total number of servers owned by cloudlet $i$. Constraints (14) and (15) indicate ${{\rho }_{i}}=\max (0,{{P}_{i}}-{{G}_{i}})$. Constraint (16) means the cloudlet provider should guarantee the SLA for all UEs. Constraint (17) implies that the total number of Avatars assigned to the cloudlet should not exceed the cloudlet's capacity and Constraint (18) means each Avatar should be assigned to no more than one cloudlet. \
	\begin{theorem}
	The problem of minimizing the on-grid energy consumption of GCN is NP-hard.
	\end{theorem} 
	
	\begin{IEEEproof}
	Suppose there are 2 cloudlets in GCN (i.e., $i=2$) and the capacity of each cloudlet is infinite (i.e., ${{m}_{1}}={{m}_{2}}=+\infty $). Meanwhile, every Avatar can migrate to any of the two cloudlets without violating Avatar propagation delay constraint (i.e., $\epsilon=+\infty$). Moreover, green energy generation of each cloudlet is the same which equals to Q (i.e., ${{G}_{1}}={{G}_{2}}=\frac{1}{2}\sum\nolimits_{k}{(\frac{{{P}_{s}}}{\tau }+\beta +\alpha {{\mu }_{k}})=Q}$) and the total energy consumption of the network is equal to the total green energy generation (i.e., $\sum\nolimits_{i=1}^{2}{{{P}_{i}}}=2Q$). So, the original problem is converted to the following:
	\begin{equation}
		\underset{\delta \left( i,k \right)}{\mathop{\min }}\,\text{   }\Delta T\sum\limits_{i=1}^{2}{\max \left\{ {{P}_{i}}-Q,0 \right\}}	
	\end{equation}
	\begin{equation}
		s.t.\text{    }\sum\limits_{i=1}^{2}{{{P}_{i}}}=2Q,
	\end{equation}
	where ${{P}_{i}}=\sum\nolimits_{k}{\{\delta (i,k)({{P}_{s}}/\tau +\beta +\alpha {{\mu }_{k}})\}}$ . Obviously, the optimal solution for minimizing the total on-grid energy consumption of GCN is to assign the total energy demands into the two cloudlets equally, i.e., $P_1=P_2=Q$, which can be considered as the partition problem (i.e., a well know NP-hard problem). So the problem of minimizing the on-grid energy consumption of the GCN is NP-hard.\
	\end{IEEEproof}
	
	To solve GEAR (which is a mixed integer linear programming problem), we use the Branch and Bound search method \cite{23} to find the sub-optimal solution to the problem. Therefore, in each time slot, each Avatar estimates its average CPU utilization for the next time slot by adopting the CPU workload prediction model \cite{24}\cite{25}, acquires the location of its UE, and reports the information to the GCN manager. The GCN manager, i.e., a central controller in GCN, decides the location of all Avatars by solving the above optimization problem. 	
\section{Simulation Result}
\begin{table}[!htb]
	\renewcommand{\arraystretch}{1.3}
	\caption{System Parameter}
	\label{tab:para}
	\centering
	\begin{tabular}{c c}
		\hline
		\hline
		Parameter  &  Value\\
	
		\hline
		
		The length of time slot, $\Delta T$   &   15 $mins$\\
		
		Capacity of server, $\tau$     &   16 Avatars\\ 
		
		Power consumption of standby server, $P^s$ & 80 $Watts$\\
		
		CPU usage to power mapping coefficient, $\alpha$ &  0.2 $\%CPU/Watts$ \\
		
		Avatar to power mapping coefficient, $\beta$		&    0.3 $Watts/Avatar$\\
	
		Distance to delay mapping coefficient, $\sigma$ & 3.33 $ms/km$\\ 
	
		SLA, $\epsilon$  &  10 $ms$\\
	
		\hline
		\hline
	\end{tabular}
\end{table}
We simulate the proposed GEAR strategy in GCN. For comparisons, we select the other Avatar migration strategy, i.e., Follow me AvataR (\emph{FAR}) migration strategy. The idea of FAR is similar to the previous work \cite{7}, which tries to minimize the propagation delay between a UE and its VM in the cloud. Similarly, FAR does not minimize the on-grid energy consumption but minimizes the propagation delay between a UE and its Avatar by selecting the nearest cloudlet as the host of the UE's Avatar (i.e., when the UE moves from one eNB coverage area to the other eNB coverage area, its Avatar also migrates correspondingly). Some system parameters are listed in Table I.\
\begin{figure}[!htb]
	\centering	
	\includegraphics[width=0.6\columnwidth]{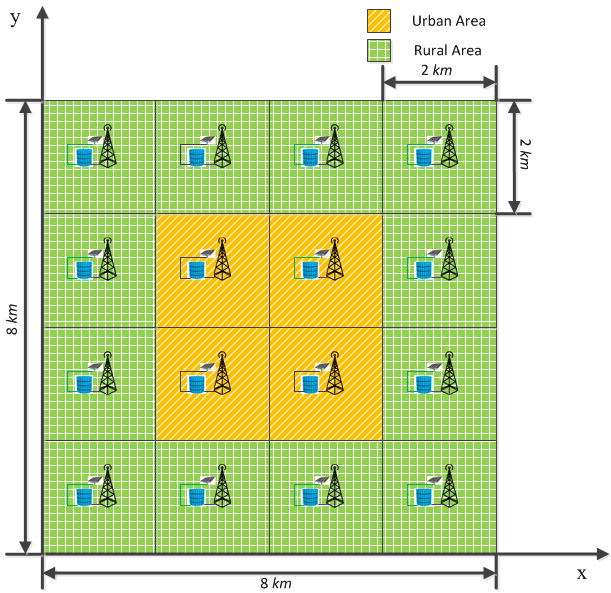}
	\caption{Network topology}
	\label{fig4}
\end{figure}
\begin{figure}[!htb]
	\centering	
	\includegraphics[width=0.6\columnwidth]{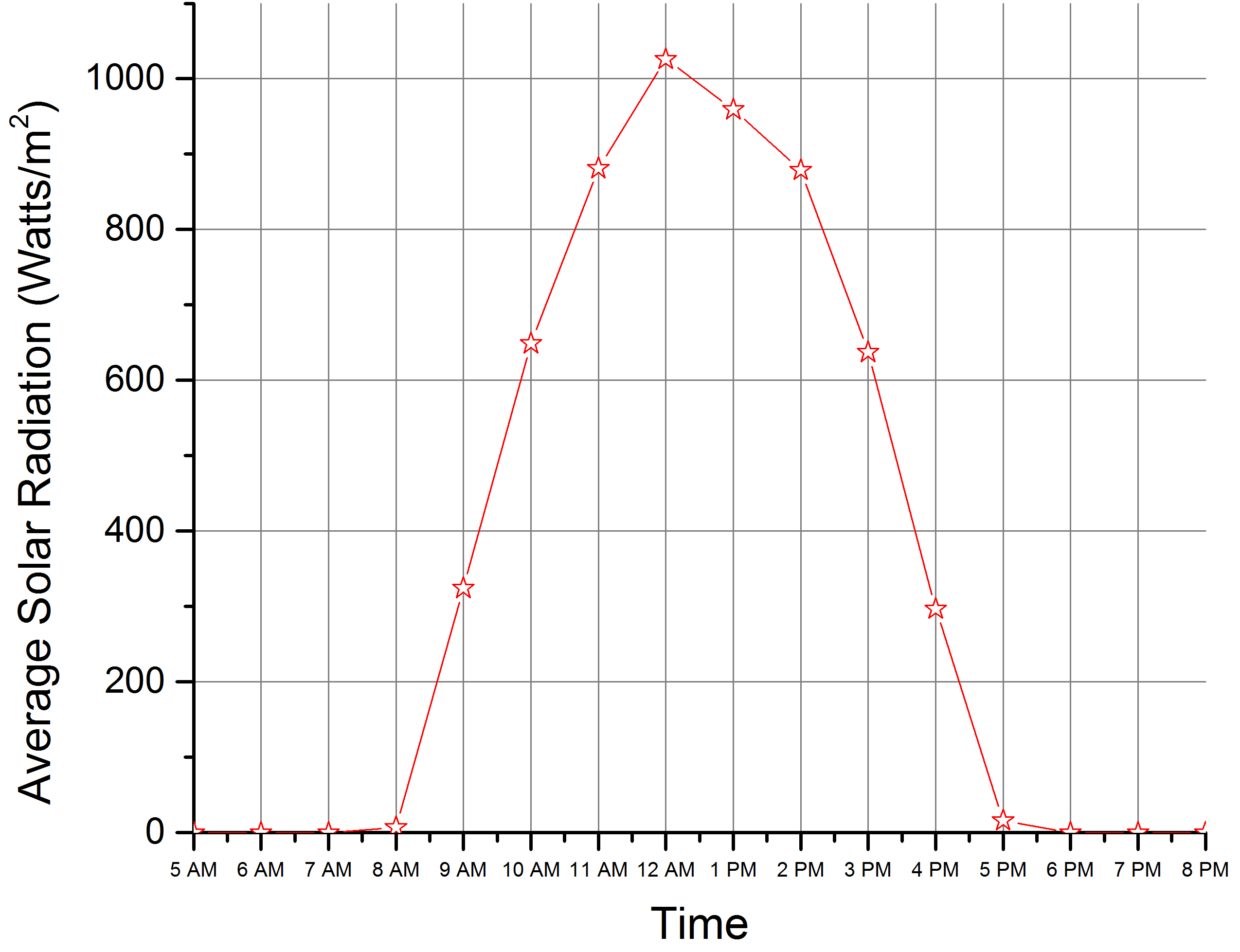}
	\caption{Average solar radiation generated at different time}
	\label{fig5}
\end{figure}
\begin{figure}[!htb]
	\centering	
	\includegraphics[width=0.6\columnwidth]{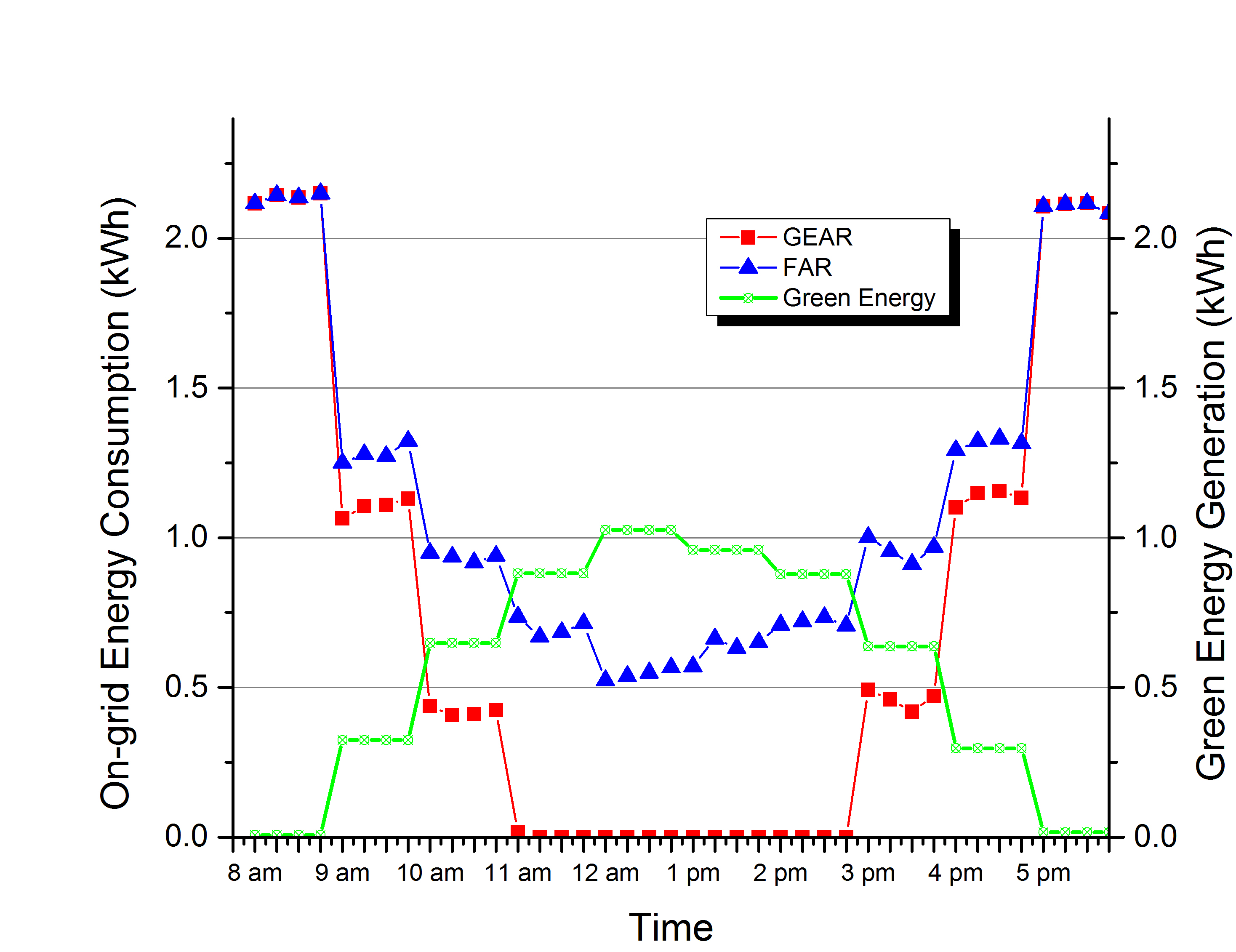}
	\caption{On-grid energy consumption at different time}
	\label{fig6}
\end{figure}
\begin{figure}[!htb]
	\centering	
	\includegraphics[width=0.6\columnwidth]{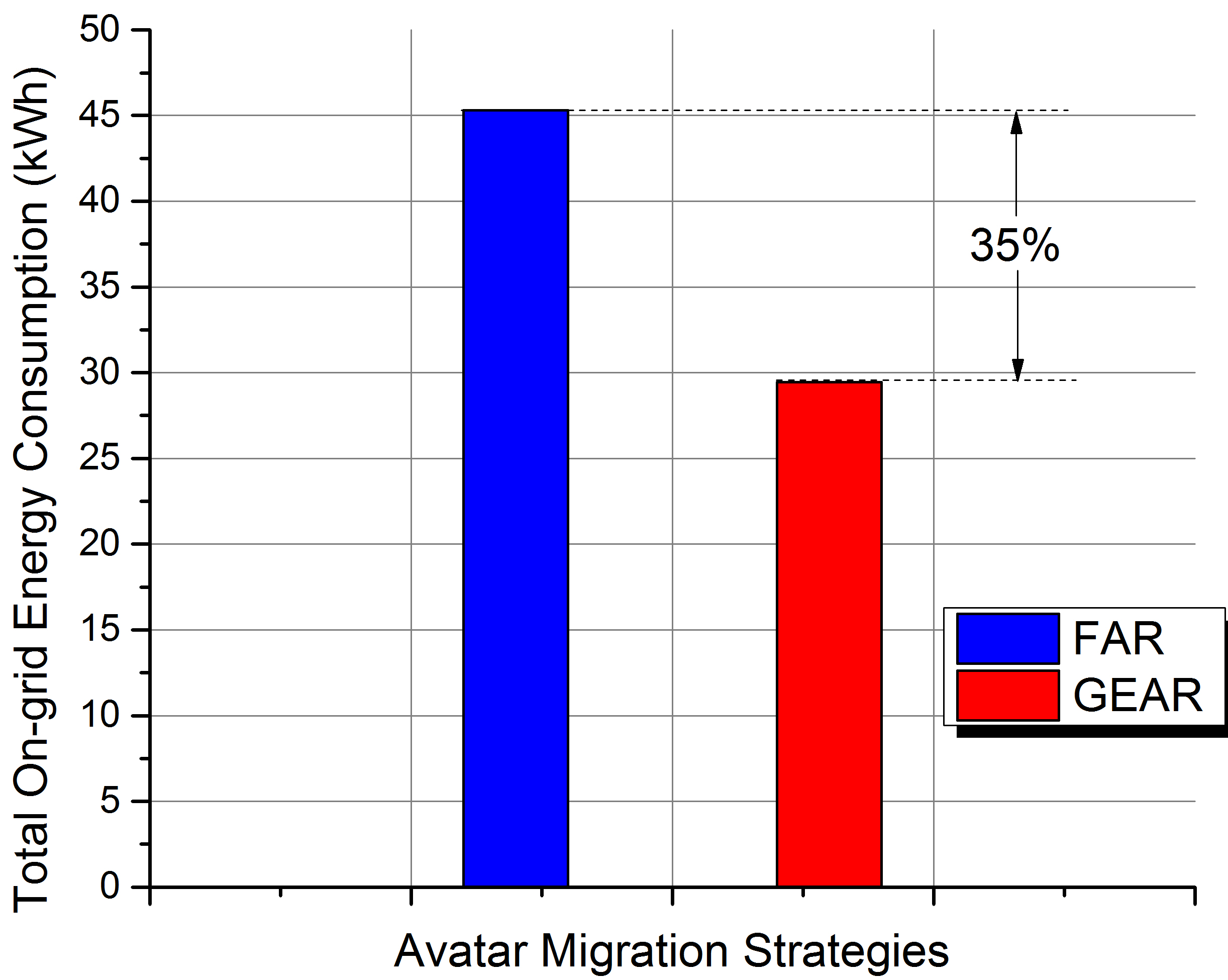}
	\caption{Total on-grid energy consumption in one day}	
	\label{fig7}
\end{figure}
\begin{figure}[!htb]
	\centering	
	\includegraphics[width=0.6\columnwidth]{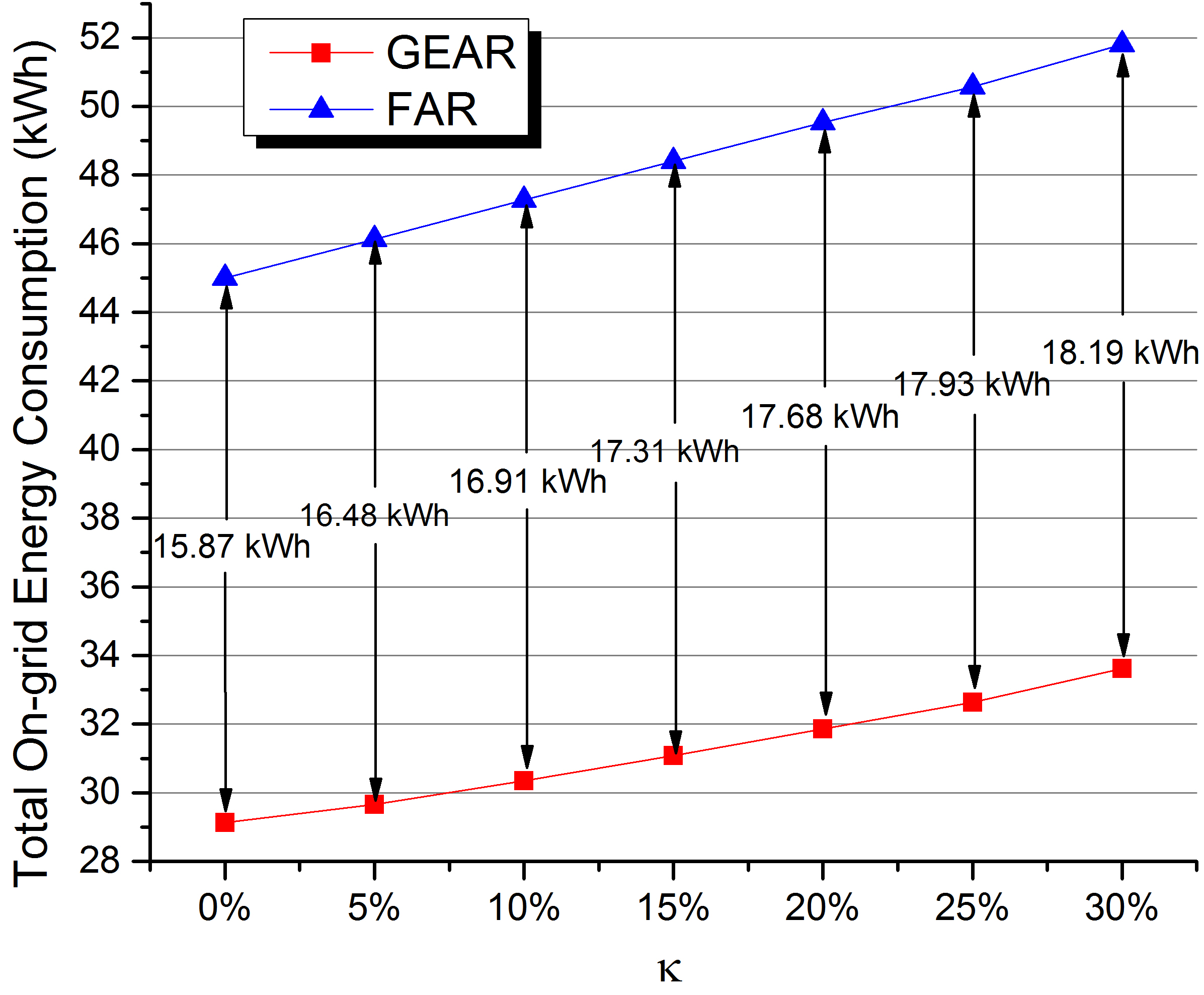}
	\caption{Total on-grid energy consumption over the  number of UEs in GCN}
	\label{fig8}
\end{figure}
\begin{figure}[!htb]
	\centering	
	\includegraphics[width=0.6\columnwidth]{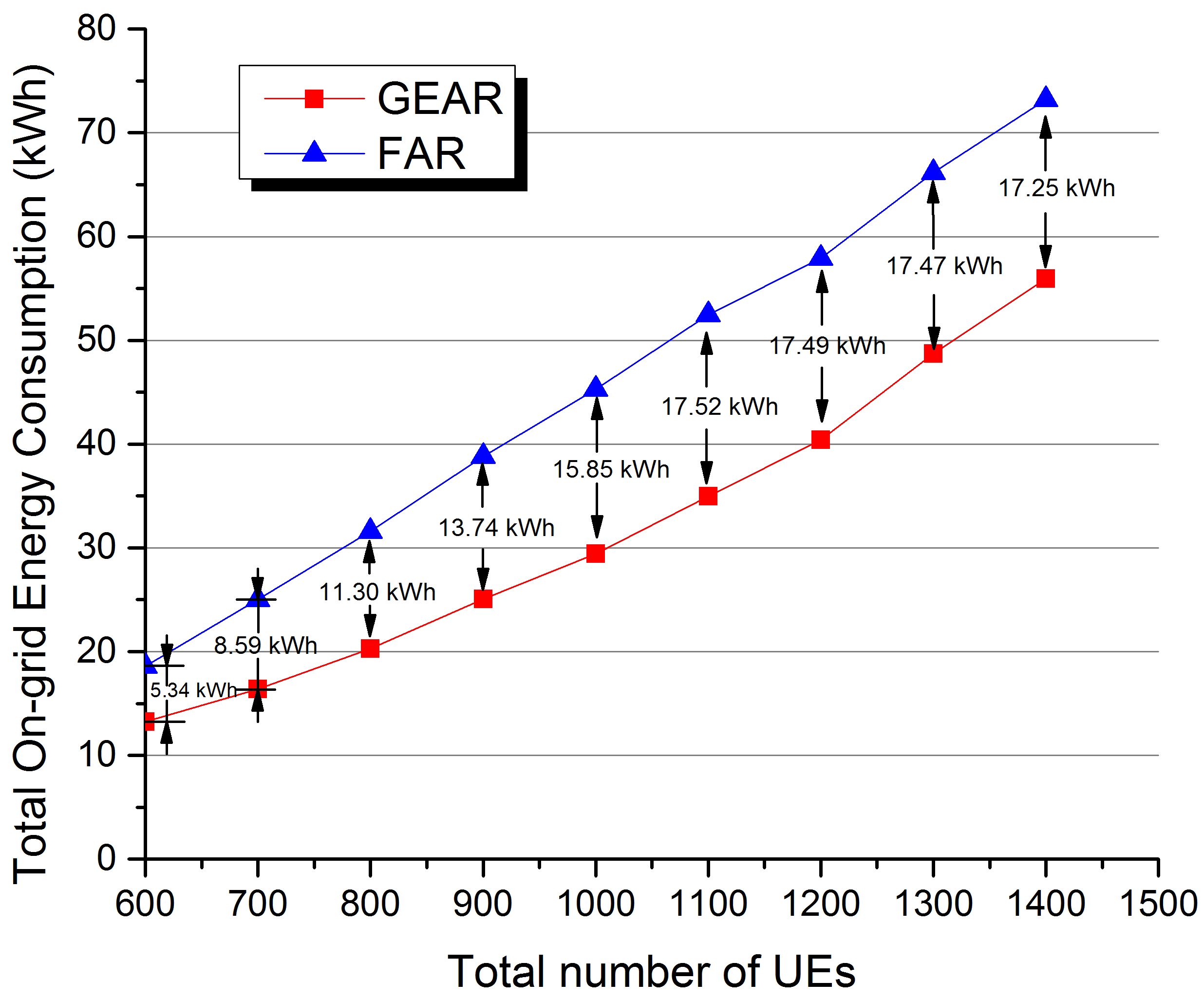}
	\caption{Total on-grid energy consumption over different values of $\kappa$}	
	\label{fig9}
\end{figure}

To demonstrate the viability of GEAR, we set up a network with the topology shown in Fig. \ref{fig4}, which includes 16 cloudlet-eNB combinations (4$\times$4) in a square area of 64 $km^2$. The coverage of each eNB is a square area of 4 $km^2$. The whole area is divided into 2 parts, i.e., urban and rural areas. Initially, each cloudlet's capacity $m_i$ in terms of the number of servers is randomly chosen between 10 and 30, and UEs are uniformly distributed in the network. The location of each Avatar is initially chosen to be its nearest cloudlet. Each Avatar's CPU is assigned by one physical core in the server and the CPU utilization of each Avatar is randomly chosen between 10\% and 100\% in each time slot (we assume the OS kernel instances cost 10\% of the Avatar's CPU utilization). Each server can host 16 Avatars at most.\
\subsection{Spatial Dynamics of Energy Demand}
Energy demands of different cloudlets in GCN exhibit spatial dynamics, and so we setup the simulation scenario as follows: UE mobility adopts the modified random waypoint model, i.e., each UE randomly selects a speed between 0 and 10 $m/s$ in every time slot and moves toward its destination, and the locations of UEs' destinations (i.e., the values of x and y coordinates) are randomly selected according to a normal distribution~N(4 $km$,1.4 $km$), which implies that UEs more likely move toward the center of each urban area (i.e., based on the characteristics of the normal distribution, UEs more likely select their destinations which are close to the center of the network). For the green energy generation, we use the local daily solar radiation data trace (Millbrook, NY in Jan. 1st, 2015) from National Climatic Data Center \cite{26}, as shown in Fig. \ref{fig5}, where each point indicates the average solar radiation within the current hour. Assume the size of the solar cell equipped in each cloudlet is 5 $m^2$ and the efficiency for converting solar radiation into electricity is 46\% \cite{27}. Also, suppose the green energy generated in different cloudlets is the same in the same time slot. Fig. \ref{fig6} shows the total on-grid energy consumption of two Avatar migration strategies in different time slots. When there is no or little green energy provision in GCN (from 8 a.m. to 9 a.m.), there is no difference between the two live migration strategies since all the cloudlets are powered by on-grid energy. However, when more green energy is generated at each cloudlet, GEAR can save more on-grid energy than FAR, since GEAR can migrate Avatars from the cloudlets with higher energy demand to the cloudlets with lower energy demand so that green energy can be fully utilized. Fig. \ref{fig7} shows the total on-grid energy consumption in GCN in the whole day; note that GEAR saves 35\% on-grid energy as compared to FAR.\

We next examine the effect of the density of UEs by increasing the number of UEs from 600 to 1400. More UEs in the network result in more out-of-balanced energy demand between the urban area and the rural area because all the UEs prefer to go to the urban rather than the rural area (according to the simulation set-up that we mentioned previously). Fig. \ref{fig8} shows the total on-grid energy consumption in one day with respect to different numbers of UEs in the network. We can see that when the number of UEs increases from 600 to 1100, the difference of the total on-grid energy consumption between GEAR and FAR is increasing, i.e., if the energy demand is more unbalanced between areas, GEAR can save more on-grid energy as compared to FAR by balancing the energy gap among the cloudlets. However, as the number of UEs exceeds 1100, the difference of the total on-grid energy consumption between GEAR and FAR remains static because green energy has already been fully utilized by GEAR when the number of UEs reaches 1100, and if the energy demands are still increasing, on-grid energy has to be tapped.\

\subsection{Spatial Dynamics of Green Energy Provision and Energy Demand}
In the real environment, not only the energy demand but also the green energy provision may exhibit spatial dynamics, i.e., the solar cell in different cloudlets may generate different amount of green energy because of the position of the sun, the spatial dynamics of atmospheric conditions, etc. Also, evidence shows that the solar radiation of the rural area is greater than the urban area \cite{28}. Therefore, we setup the simulation scenario as follows: the UE's parameters are the same as in the previous simulation scenario and the hourly average solar radiation in the rural area still follows the data trace as shown in Fig. \ref{fig5}. However, the hourly average solar radiation in the urban area is reduced by κ percentage (the value of κ is selected from 0\% to 30\% in the simulation). Fig. \ref{fig9} shows the total on-grid energy consumption of the network in one day with respect to different values of $\kappa$. Note that as the value of $\kappa$ increases (i.e., hourly green energy generation is getting less in the urban area and the energy gap between the two areas is getting larger), FAR consumes more on-grid energy because the energy gap of the urban area is getting larger, and GEAR can save more on-grid energy as compared to FAR by balancing the energy gap between the two areas.

\section{Conclusion}
In this paper, we have proposed the GCN architecture to provide seamless and lower latency MCC services to UEs, i.e., UEs can offload tasks to their powerful Avatars with shorter propagation delay. However, owing to the spatial dynamics of energy demand and green energy provisioning, a significant amount of green energy is wasted, thus resulting in more grid energy consumption. Therefore, we have proposed the GEAR strategy to redistribute the energy demand by migrating Avatars among cloudlets according to cloudlets' green energy generation and to guarantee the maximum Avatar propagation delay. Simulation results have demonstrated that GEAR can significantly save on-grid energy as compared to the FAR strategy.\

In the future, we will consider the heterogeneous nature of the cloudlets, i.e., the configurations of the servers in GCN are different, and UEs can choose different configurations of VMs as their Avatars. Also, Avatar migration cost will be considered in the optimal Avatar migration strategy. Moreover, we will also consider the scenario in which a cloudlet and an eNB in the same location can share green energy and both of them can also generate different energy demands in different time slots. It is a big challenge to design an optimal Avatar migration strategy by considering the energy demands of eNBs and cloudlets.

\bibliographystyle{IEEETran}

\end{document}